\documentclass[10pt,a4paper]{iopart}

\usepackage{iopams, gensymb, color}
\usepackage[pdftex]{graphicx}
\DeclareGraphicsExtensions{.pdf,.jpg}
\usepackage{epstopdf}

\definecolor{myRed}{rgb}{0.8,0.3,0.3}
\definecolor{myGreen}{rgb}{0,0.7,0.5}
\definecolor{myBlue}{rgb}{0.2,0.4,0.9}
\definecolor{myPurple}{rgb}{0.3,0.3,0.8}
\definecolor{myOrange}{rgb}{0.9,0.5,0.2}

\newcommand{\micron}{\mu\mathrm{m}}
\newcommand{\MeV}{\mathrm{MeV}}
\newcommand{\Wcm}{\mathrm{Wcm}}
\newcommand{\fs}{\mathrm{fs}}
\newcommand{\Energy}{\mathcal{E}}

\begin{document}

\title{Measuring quantum radiation reaction in laser--electron-beam collisions}

\author{T G Blackburn}
\address{Clarendon Laboratory, University of Oxford, Parks Road, Oxford, OX1 3PU, UK}
\ead{tom.blackburn@physics.ox.ac.uk}
\vspace{10pt}


\begin{abstract}
Today's high-intensity laser facilities produce short pulses can, in tight focus,
reach peak intensities of $10^{22}~\Wcm^{-2}$ and, in long focus, wakefield-accelerate
electrons to GeV energies.
The radiation-reaction--dominated regime, where the recoil from stochastic photon emission
becomes significant, can be reached in the collision of such an electron beam
with an intense short pulse. Measuring the total energy emitted in gamma rays or
the presence of a prominent depletion zone in the electron beam's post-collision energy spectrum
would provide strong evidence of radiation reaction, provided enough electrons penetrate
the region of highest laser intensity. Constraints on the accuracy of timing necessary
to achieve this are given for a head-on collision.
\end{abstract}


\vspace{2pc}

\submitto{\PPCF}
%
%
%


\section{Introduction}
\label{sec:Introduction}

	\begin{figure}
	\centering
	\makeatletter
	\if@twocolumn
		\includegraphics[width=0.8\linewidth]{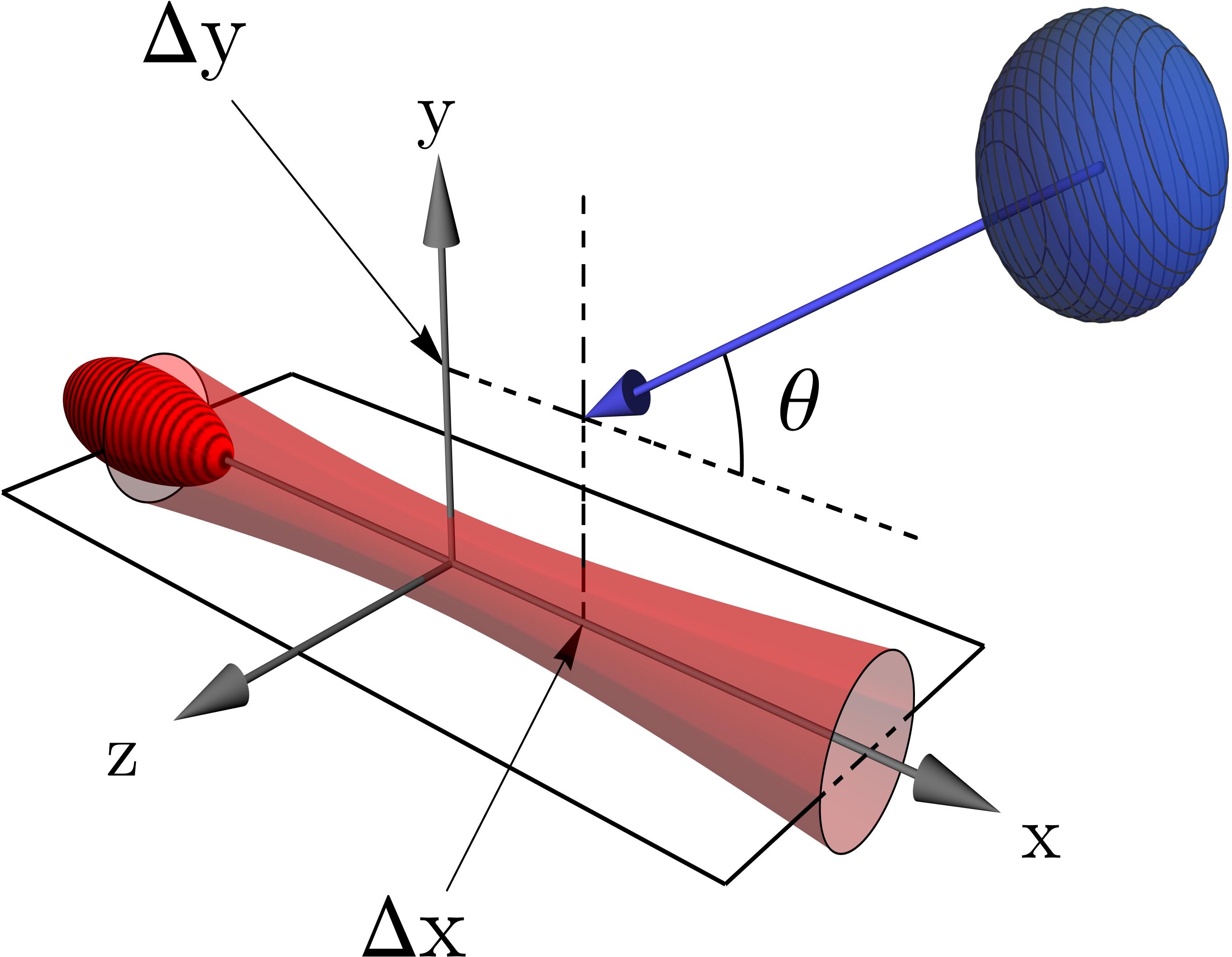}
	\else
		\includegraphics[width=0.5\linewidth]{figure1.jpg}
	\fi
	\makeatother
	\caption{The experimental configuration under consideration: a GeV electron beam (blue)
			colliding with an laser pulse of intensity $10^{22}~\Wcm^{-2}$ (red). The laser
			moves in the positive $x$-direction; the electron beam centre is located in,
			and the electrons' initial momenta are parallel to, the $x$-$y$ plane.
			Observing radiation reaction will depend upon the degree to which the beams
			overlap, i.e. the collision timing and angle $\theta$. The timing error is parameterised by
			$(\Delta x, \Delta y)$, the location of the electron beam centre at $t=0$, when
			the laser pulse is focussed at the origin. We do not consider the effects of a
			displacement between the beams in the $z$-direction.}
	\label{fig:diagram}
	\end{figure}

The development of multi-PW laser systems, and the prospect of focussing laser pulses to intensities
greater than $10^{22}~\Wcm^{-2}$~\cite{vulcan,eli}, has led to substantial interest in how
radiation reaction and strong-field quantum electrodynamics (QED) processes will alter the
plasma physics studied over the next decade.
Future laser systems will be powerful enough such that the
high fluxes of gamma rays (i.e. photons in
the MeV range) and electron-positron
pairs they produce in laser--laser~\cite{bell,kirk,duclous} and
laser--solid~\cite{ridgers2012,brady} interactions
exert significant feedback on the evolution of the plasma, increasing the
absorption of laser energy~\cite{nerush}.
It will become possible to test the physics
underpinning exotic astrophysical phenomena, e.g. pair cascades in pulsar magnetospheres~\cite{goldreich,timokhin}, and illuminate fundamental questions about radiation
reaction and the quantum vacuum in QED~\cite{diPiazza2012}.

The importance of strong-field QED effects may be quantified with the
parameter
	\begin{eqnarray}
	\eta = \frac{|F_{\mu\nu}p^{\nu}|}{m c E_\mathrm{Sch}}
	\label{eq:EtaDef}
	\end{eqnarray}
where $F_{\mu\nu}$ is the electromagnetic field tensor and $p^{\mu} = \gamma m(c,\mathbf{v})$
the electron four-momentum~\cite{bell}. (This parameter is also called $\chi$~\cite{mackenroth}
or $\chi_e$~\cite{bulanov}.)
$\eta$ compares the magnitude of the electric field in the electron's rest frame to that of
the critical field of QED $E_\mathrm{Sch}=
1.3\times10^{16}~\mathrm{Vcm}^{-1}$, which can produce electron-positron pairs
directly from the vacuum~\cite{schwinger}.

$\eta$ also controls the stochasticity of photon emission. As the typical energy of a radiated
photon for $\eta<1$ is $\hbar \omega \simeq 0.44 \eta \gamma m c^2$~\cite{bell},
when $\eta \simeq 1$, an electron can lose a substantial fraction of its energy in a single
emission. Radiation reaction must be then treated as stochastic, rather than as the continuous
loss of energy predicted classically. Pair production can become
significant in laser--solid interactions~\cite{ridgers2012,brady} for $\eta\sim1$; in a colliding beams
experiment, where the electron is not reaccelerated by the laser, $\eta \gg 1$ is required for
these processes to be significant~\cite{bulanov, nerush}.

For highly relativistic electrons, \eref{eq:EtaDef} becomes
	\begin{eqnarray}
	\eta \simeq \frac{\gamma |\mathbf{E}_\perp + \mathbf{v} \times \mathbf{B}|}{E_\mathrm{Sch}}
	\label{eq:EtaDef2}
	\end{eqnarray}
where $\gamma$ is the electron Lorentz factor, $\mathbf{B}$ the magnetic field and $\mathbf{E}_\perp$
the electric field component perpendicular to $\mathbf{v}$. In a laser-plasma interaction
we would expect the electrons to have $\gamma \simeq a_0$, where
$a_0 = [I (\lambda/\micron)^2 / 1.37\times10^{18}~\Wcm^{-2}]^{1/2}$ is the laser's
strength parameter, $I$ its intensity and $\lambda$ its wavelength. Then the typical
$\eta \simeq I\lambda/(5.65\times10^{23}~\Wcm^{-2}\micron)$ and intensities $>10^{23}~\Wcm^{-2}$
would be necessary to observe strong-field QED effects.

However, if the electrons are pre-accelerated to high energies before encountering the high
field region, the QED-dominated regime can be reached using much lower laser intensities.
This was exploited in experiment E-144 at the SLAC facility~\cite{bula,burke}, in which
the collision of a 46.6~GeV electron
beam and laser pulse of intensity $10^{18}~\Wcm^{-2}$ was observed to produce
electron-positron pairs by photon--photon scattering.
The advent of wakefield-accelerated electron beams with energies of 1 GeV~\cite{kneip,kim}
or greater~\cite{wang,leemans} and intense
short pulses~\cite{hercules} now raises
the possibility that this experiment could be repeated with an all-optical setup in
today's laser facilities.
This has already been accomplished for target laser pulses with
$a_0 \simeq 1$~\cite{chen,powers}; recently Sarri~\etal~\cite{sarri} demonstrated
production of multi-MeV gamma rays by non-linear Thomson scattering of a ${\sim}400~\MeV$
electron beam. If high-$Z$ solid rather than a laser pulse is used as a target, 
wakefield-accelerated electron beams can also be used to generate
dense electron-positron plasmas by bremsstrahlung~\cite{sarriNatComm}.

For collisions at $a_0 \simeq 1$, the electron recoil is negligible.
However, for a GeV electron beam and laser pulse of intensity $10^{22}~\Wcm^{-2}$,
$\eta \simeq 1$ and quantum radiation reaction will dominate the dynamics of the electron
beam. There is now a large body of work considering how strong-field QED effects can be
observed in the spectra of the emitted gamma rays and the electron beam of such an
experiment~\cite{thomas,neitz2013,mackenroth2013,neitz2014,ilderton,harvey,li}.

In previous work~\cite{me}, we considered the collision of a GeV electron beam 
with a laser pulse of intensity $10^{21}$ to $3\times10^{22}~\Wcm^{-2}$; the beams collided
at an angle of $180\degree$ and were timed to coincide perfectly at the laser focus.
Quantum radiation reaction could be experimentally diagnosed by measuring the increased
yield of high-energy gamma rays or the reduced energy of the electron
beam after the collision.

Both these signatures are sensitive to the overlap between the electron beam and laser pulse,
which places constraints on the accuracy of collision timing required.
In this article we will consider how the collision timing, and the angle between the beams, affects the
total gamma ray energy, the electron beam's final energy spectrum and therefore the viability of diagnosing quantum
radiation reaction in an experiment using today's high intensity laser facilities.

\Sref{sec:Method} describes the algorithm by which these interactions are simulated;
\sref{sec:Parameters} describes the chosen experimental parameters of GeV electron beam
and $10^{22}~\Wcm^{-2}$, 30~fs laser pulse.
\Sref{sec:GRProduction} presents simulated gamma ray spectra and shows that the
yield is maximised in a head-on collision that occurs $80~\micron$ from
the laser focal spot.
\Sref{sec:ElectronBeam} presents simulated electron spectra and shows that, if the collision
is accurate enough, radiation reaction leads to a substantial energy loss in that part of the
electron beam the laser passes through.
As measuring that loss can be a robust signature of radiation reaction, we constrain how
accurate a collision must be for it to be sufficiently large.


\section{Method of simulation}
\label{sec:Method}

The collision of a GeV electron beam with a laser pulse of intensity $10^{22}~\Wcm^{-2}$
is simulated with a single-particle, Monte-Carlo code that includes the electron's oscillatory
motion in the laser fields and the strong-field QED process of synchrotron photon emission.
As pair production by the resulting gamma rays is likely to be negligible for these
parameters~\cite{me}, we can neglect it here.

Following~\cite{erber, bks, ritus}, the differential rate of photon emission is
	\begin{eqnarray}
	\frac{\rmd^2 N}{\rmd t \rmd \chi} =
		\frac{\sqrt{3}\alpha}{2 \pi \tau_\mathrm{C}}
		\frac{\eta}{\gamma}
		\frac{F(\eta, \chi)}{\chi}.
	\label{eq:MasterRate}
	\end{eqnarray}
$\chi=\hbar|F_{\mu\nu}k^{\nu}|/2mcE_\mathrm{Sch}$, where the four-frequency
$k^{\mu}=(\omega/c, \mathbf{k})$, is the equivalent of $\eta$ for the photon;
the Compton time $\tau_\mathrm{C} = \hbar / m c^2$ and the quantum synchrotron function,
discussed in Appendix~A, is defined for $0 \leq \chi \leq \eta/2$.
Integrating \eref{eq:MasterRate} over that range gives the total rate of
emission $\rmd N/\rmd t \propto \eta h(\eta)/\gamma$, where the function
	\begin{eqnarray}
	h(\eta) = \int_0^{\eta/2} \! \frac{F(\eta,\chi)}{\chi} \, \rmd \chi
	\end{eqnarray}
is a monotonically decreasing function of $\eta$ that satisfies
$\lim_{\eta \rightarrow 0} h(\eta) = 5 \pi/3$.

This rate is calculated in the Furry picture of QED~\cite{furry}, including an external,
unquantised electromagnetic field. The basis states derived for this field
may be used to calculate Feynman rules for quantised interactions~\cite{seipt, heinzl, mack}. 

Where the photon formation length is much smaller the characteristic length of the fields,
that emission may be treated as pointlike and the fields quasistatically;
thus the rate can be calculated
in an equivalent system that has the same value of $\eta$.
As the photon formation length $L_\mathrm{ph} = \lambda / a_0$~\cite{kirk},
we may treat the laser fields as quasistatic provided $a_0 \gg 1$.
In Erber~\cite{erber}, the equivalent system used to derive \eref{eq:MasterRate} is a
static magnetic field in the limit $B \rightarrow 0$, $\gamma \rightarrow \infty$.

The emission algorithm follows the work of~\cite{duclous,ridgers2012,me,ridgers2014}
and is similar to that described in~\cite{harvey,elkina}.
Photon emission is a stochastic process and governed by Poisson statistics;
the probability $P$ of emission in a field with optical depth $\tau$ is $P = 1 -
e^{-\tau}$. At the start of the simulation, and following each emission, the
electron is assigned a `final' optical depth $\tau_\mathrm{f}
= -\log (1 - P)$ for pseudorandom  $P \in [0,1]$. The electron's optical depth $\tau$ is
integrated along its trajectory until $\tau_\mathrm{f}$ is reached, according to
	\begin{eqnarray}
	\frac{\rmd \tau}{\rmd t}
		= \int_0^{\eta/2} \! \frac{\rmd^2 N}{\rmd t \rmd \chi} \, \rmd \chi
		= \frac{\sqrt{3}\alpha}{2 \pi \tau_\mathrm{C}} \frac{\eta}{\gamma} h (\eta).
	\label{eq:TauRate}
	\end{eqnarray}
At this point, the photon energy $\hbar \omega = 2 \chi m c^2 /\eta $ is selected by solving
	\begin{eqnarray}
	r\,h(\eta) = \int_0^\chi \! \frac{F(\eta, \chi')}{\chi'} \, \rmd \chi'
	\end{eqnarray}
for $\chi$, where the pseudorandom number $r$ is
uniformly distributed in $[0,1]$. The electron then recoils antiparallel to its motion,
with change of momentum $\hbar \omega/c$. Energy conservation therefore requires that a small amount
of energy is transferred from the laser fields during emission; as described in~\cite{duclous},
this leads to a fractional error in the electron energy of $\Delta \gamma / \gamma \propto 1/\gamma$
that may be safely neglected. While synchrotron radiation is directed forward into a cone of
opening angle ${\sim}1/\gamma$~\cite{landau}, we can assume that the photon and electron momenta
are collinear, as the electron is always highly relativistic.

Between emissions, the electron trajectory is determined classically by integrating the Lorentz
force exerted by the laser fields (neglecting the beam's space charge field, as described in
\sref{subsec:ElectronBeam})
	\begin{eqnarray}
	\frac{\rmd \mathbf{r}}{\rmd t} = \frac{\mathbf{p}}{\gamma m}
	\\
	\frac{\rmd \mathbf{p}}{\rmd t} = -e (\mathbf{E} + \mathbf{v} \times \mathbf{B}).
	\end{eqnarray}
This is implemented using a Boris push algorithm~\cite{boris}, so $\mathbf{E}$ and
$\mathbf{B}$ are calculated at intervals staggered by half a timestep. 

The electron trajectories are discretised into timesteps of $0.005~\fs$, and at least
$10^{7}$ trajectories are followed for each set of parameters.


\section{Parameters}
\label{sec:Parameters}

A schematic of the experimental configuration under consideration is given in \fref{fig:diagram}.
We consider varying the angle of collision $\theta$ as well as the mistiming in
the $x$-$y$ plane. This is parameterised by the offsets $(\Delta x, \Delta y)$, which correspond to the
displacement of the electron beam centre from the origin at $t = 0$, at which time the laser
is focussed there.


\subsection{Laser pulse}
\label{subsec:LaserPulse}

The laser is a Gaussian focussed beam with waist $w_0 = 2~\micron$ and wavelength $\lambda =
0.8~\micron$, linearly polarised along the $y$ axis and propagating towards $+x$.
It has a Gaussian temporal intensity profile with maximum $I = 10^{22}~\Wcm^{-2}$ ($a_0 = 65$)
and full length at half maximum $f = c \cdot 30~\fs = 9~\micron$.
The peak field $E_0$ is related to the laser intensity by $I = \frac{1}{2}c \epsilon_0 E_0^2$.

It can be shown from \eref{eq:EtaDef2} that $\eta$, which controls photon emission, does not
depend on the polarisation of the EM field, but only on the angle between the optical axis
and the electron momentum $\theta$. As such, the choice of polarisation here is arbitrary
except that it defines the plane in which the electrons oscillate and so affects
the angular distribution of radiation. This will be a fan of opening angle
${\sim}a_0/\gamma$ oriented along the polarisation axis~\cite{me}.


\subsection{Electron beam}
\label{subsec:ElectronBeam}

The energy distribution of the electron beam (up to a normalising constant) is
	\begin{equation}
	\frac{\rmd N}{\rmd \Energy} \propto (\mu + \sigma/3 - \Energy)^{-3/2}
		\exp \left( -\frac{\sigma}{2 (\mu + \sigma/3 - \Energy)} \right)
	\end{equation}
for $100~\MeV \leq \Energy \leq \mu + \sigma/3$; it has a peak at
$\mu = 1000~\MeV$ and a width $\sigma$ of $250~\MeV$, which is related to the full width
at half maximum (fwhm) $f$ by $f = 0.9 \sigma$. This is shown in \fref{fig:ExamplePlot}.
This distribution has been chosen as high energy
wakefield-accelerated beams typically have large tails that extend to low energy.
They also exhibit an angular divergence of order mrad,
but here their initial momenta are directed along $\mathbf{\hat{p}}_0=-(\cos\theta,\sin\theta,0)$.

The charge density of the beam forms a Gaussian envelope with
$\sigma_\parallel = 3.8~\micron$ and $\sigma_\perp = 4.2~\micron$
($\mathrm{fwhm}_\parallel = 9~\micron$ and $\mathrm{fwhm}_\perp = 10~\micron$) where the widths
are given in the directions parallel and perpendicular to $\mathbf{\hat{p}}_0$ respectively.

The electric and magnetic components of the space charge force
$F_\mathrm{sc} \simeq Qe / (\gamma^2 \epsilon_0 \sigma^2_\perp)$ of
an electron bunch that has cylindrical symmetry nearly cancel if
the electrons are ultrarelativistic $\gamma \gg 1$.
The force of radiation reaction $F_\mathrm{rr} \simeq \alpha \eta^2 m c / \tau_\mathrm{C}$~\cite{bell}.
For $Q \simeq 100~\mathrm{pC}$, $\sigma_\perp \simeq 4~\micron$ and $\eta \simeq 0.1$,
$F_\mathrm{sc} \ll F_\mathrm{rr}$ means that
the effect of the space charge field can be neglected and each electron's
trajectory in the laser pulse may be considered separately.

Furthermore, as the typical separation between electrons $(\sigma_\parallel \sigma_\perp^2 e /Q)^{1/3}
\simeq 5~\mathrm{nm}$ (equivalent to 40~eV), we can also neglect collective radiation effects
when considering gamma ray emission.


\section{Gamma ray production}
\label{sec:GRProduction}

	\begin{figure*}
	\centering
	\includegraphics[width=0.45\linewidth]{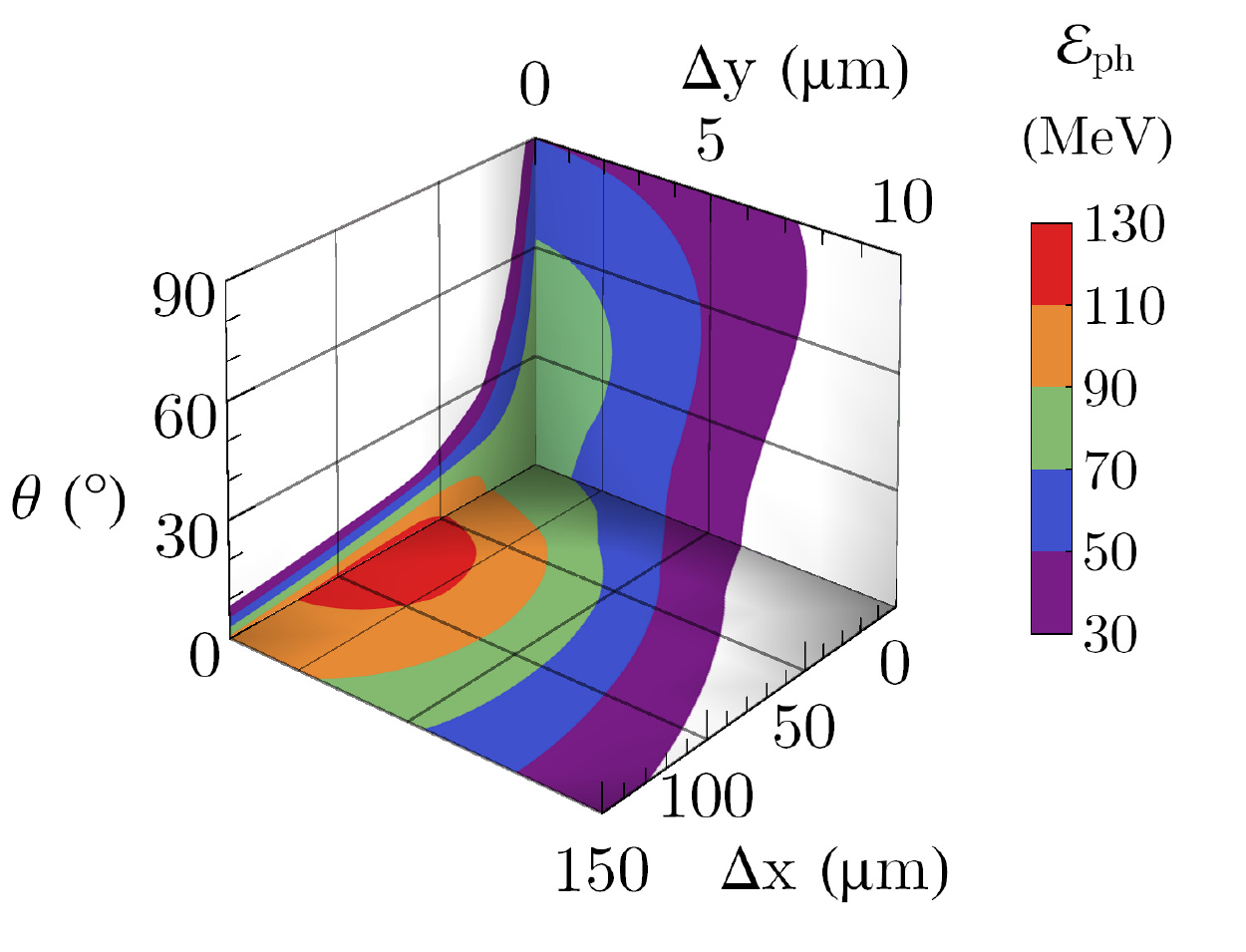}
	\includegraphics[width=0.45\linewidth]{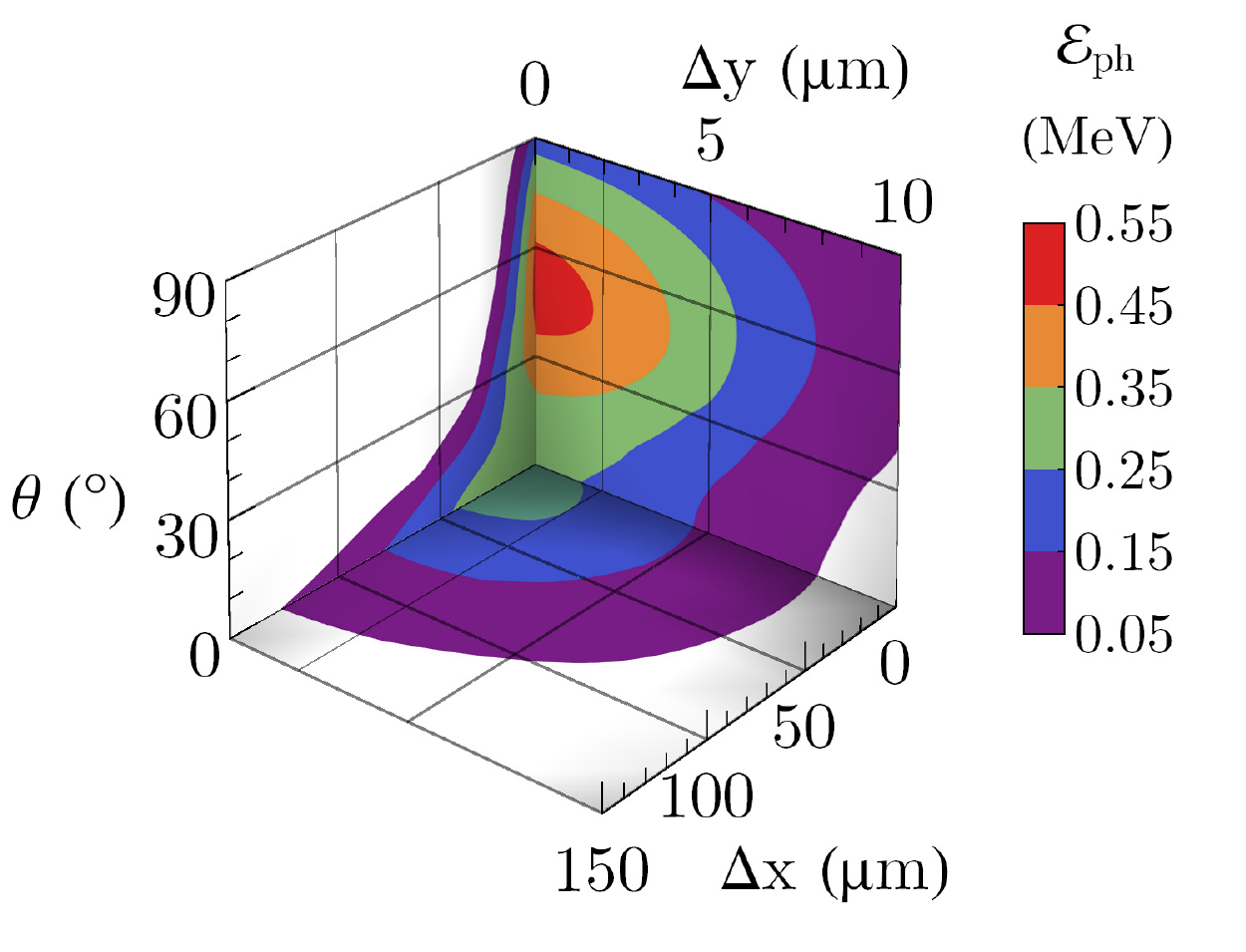}
	\caption{The energy radiated by a GeV electron beam, per electron, to photons of all energies
			(left) and to photons with $\hbar \omega > 500~\MeV$ (right), when colliding
			with a Gaussian laser pulse of intensity $10^{22}~\Wcm^{-2}$ at an
			angle $\theta$ with parallel and perpendicular offsets $\Delta x$ and $\Delta y$.}
	\label{fig:ElectronBeam}
	\end{figure*}
	
We consider the interaction of an intense laser pulse with a GeV electron beam,
with parameters as described in \sref{subsec:ElectronBeam}.
\Fref{fig:ElectronBeam} shows the mean energy lost by such an electron beam to
gamma rays when colliding with the specified laser pulse.

We find that the total loss reaches its maximum of 120~MeV per electron for a head-on collision
that occurs a distance $\Delta x = 80~\micron$ along the optical axis from the laser focus,
but that the loss to photons with energy $> 500~\MeV$ is maximised at 0.48~MeV per electron for a
perfectly-timed collision that occurs at an angle of $50\degree$.

In determining the $(\Delta x, \Delta y, \theta)$ that maximise the energy emitted to gamma rays,
we must consider three competing factors: a) the geometric factor in $\eta$; b) each electron's
length of interaction with the laser pulse; and c) the overlap between the electron beam and
laser pulse.

a) arises from the definition of $\eta$ \eref{eq:EtaDef2}: if
$\mathbf{E}$ and $\mathbf{B}$ are given by a linearly polarised plane wave and the electron
is highly relativistic, travelling at angle $\theta$ to the wave's direction of propagation,
then
	\begin{eqnarray}
	\eta = \frac{\gamma |\mathbf{E}_\perp + \mathbf{v} \times \mathbf{B}|}{E_\mathrm{Sch}}
		= \frac {\gamma (1 + \cos \theta) |\mathbf{E}|}{E_\mathrm{Sch}}.
	\end{eqnarray}
As $\eta$ falls with increasing $\theta$, colliding the electron with the laser pulse at an
angle will have two effects: according to \eref{eq:TauRate} the electron's emission rate
is reduced; and the photons it emits have lower energies, as the tail of the synchrotron
spectrum grows non-linearly with increasing $\eta$.

b) accounts for the importance of `straggling'~\cite{duclous}: as photon emission is
probabilistic, some electrons can reach the centre of the laser pulse having lost no energy;
their $\eta$ is then much higher and so too their probability of emitting a single photon with
$\hbar \omega \sim \gamma m c^2$. In this case, the laser pulse is longer along
the propagation axis than it is wide, so electrons traversing the pulse at an angle interact with
the high intensity field over a shorter distance. While this reduces the total energy emitted,
straggling becomes likelier with a shorter distance to reach the region of highest intensity;
this should increase the yield of the highest energy photons.

We can see the interplay of these two factors in the yield of the highest
energy gamma rays, which is maximised for a collision at $50\degree$ rather than at $0\degree$.
This can be explained by estimating analytically the average $\eta$ of an electron
when it emits its first photon, which balances the reduced geometric factor against
increased penetration into the laser pulse as $\theta$ increases. The latter is likely
to be more important than both the geometric factor and the overlap between the beams as 
it is the emission of the highest energy photons that
is most sensitive to straggling~\cite{me}. The calculation is given
in Appendix~B; for the parameters under consideration here, $\eta$ is maximised for a
collision $\theta = 65\degree$, which is consistent with the peak shown in the right of
\fref{fig:ElectronBeam}.
We conclude that it is most important to reduce
the length of interaction and so the energy loss in the foot of the pulse
to maximise the yield of the highest energy photons.

	\begin{figure}
	\centering
	\makeatletter
	\if@twocolumn
		\includegraphics[width=\linewidth]{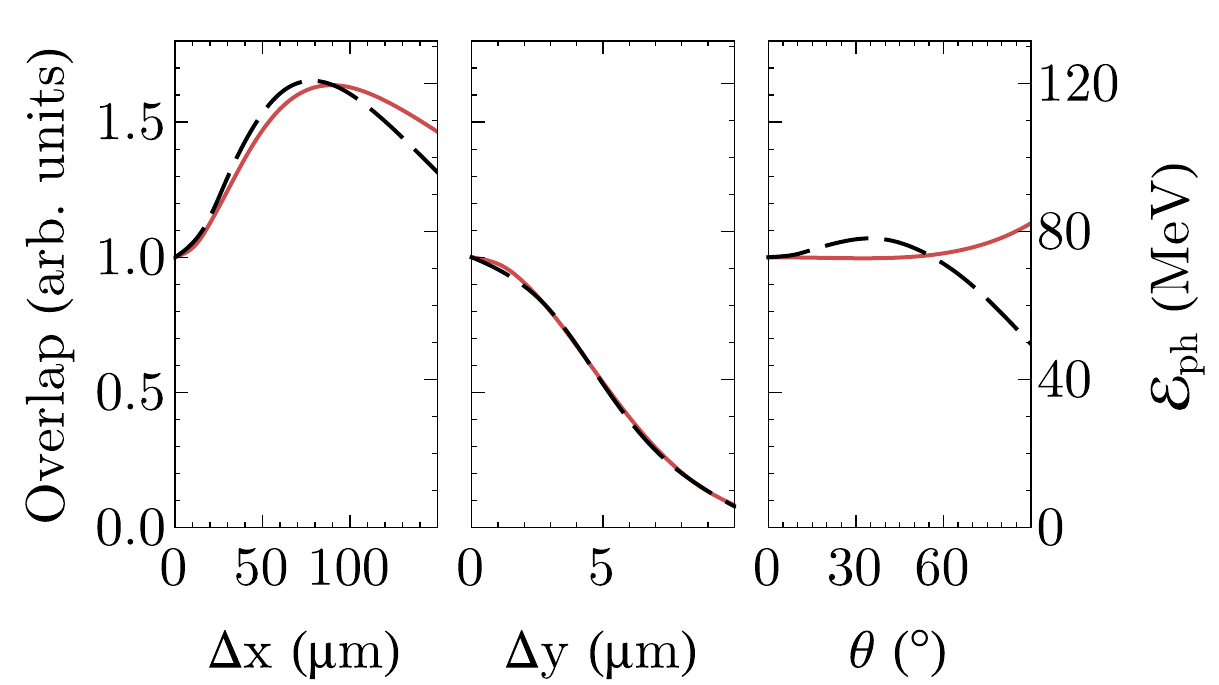}
	\else
		\includegraphics[width=0.6\linewidth]{figure3}
	\fi
	\makeatother
	\caption{The overlap between the laser pulse and electron beam (red, solid: left scale),
			normalised to the overlap at $\Delta x = \Delta y = \theta = 0$, and the 
			total loss of energy to photons per electron (black, dashed: right scale),
			varying only $\Delta x$ (left), $\Delta y$ (centre) and $\theta$ (right).}
	\label{fig:BeamOverlap}
	\end{figure}

However, maximising the total yield of gamma rays is achieved for a head-on collision at
$\Delta x = 80\micron$. We can understand this in terms of c), the overlap, which controls
how many of the beam electrons interact with the laser pulse. Let us compare the
total energy emitted in photons to an analytical estimate of the overlap
	\begin{eqnarray}
	\Omega = \int \! \rmd t \int \! \rmd^3 \mathbf{r} \,
		\rho ( \mathbf{r}, t ) |E (\mathbf{r}, t )|
	\label{eq:Overlap}
	\end{eqnarray}
where the beam charge density is as given in \sref{subsec:ElectronBeam} and
$| E (\mathbf{r},t ) |$ is as given in \sref{subsec:LaserPulse}, taking
$|\sin\phi| = 2/\pi$. That comparison is plotted in \fref{fig:BeamOverlap}, varying each of
$\Delta x$, $\Delta y$ and $\theta$.

We can see that the overlap models the dependence of the yield on $\Delta x$ and
$\Delta y$ well. The yield increases with $\Delta x$ up to $\Delta x = 80~\micron$ because the
laser pulse diverges as it propagates away from its focal plane and therefore more
electrons encounter the pulse. For $\Delta x > 80~\micron$ the yield falls because
even though more electrons collide with the laser, they do so at lower peak intensity,
which reduces their $\eta$ and so their emissivity.

Gamma ray production is more sensitive to the perpendicular displacement between the
beams $\Delta y$ than the parallel displacement $\Delta x$. This is because the peak intensity
at the electron beam centre falls as $\exp(-2(\Delta y/w_0)^2)$ for increasing $\Delta y$ but
as $[1+(\Delta x/x_\mathrm{R})^2]^{-1}$ for increasing $\Delta x$; to reduce the peak intensity
by a half requires $\Delta x = 16~\micron$ but only $\Delta y = 1.4~\micron$. For
$\Delta y > 10~\micron$ there is negligible photon emission because nearly all the
electrons have missed the laser pulse.

The overlap between the beams is nearly constant if only $\theta$ is varied; however,
the gamma ray yield is maximised at 80~MeV per electron at $\theta = 35\degree$ and
falls thereafter. We can attribute this behaviour to the reduced length of interaction
at large $\theta$, which reduces the energy loss before the electron reaches the
pulse centre. However, this does not increase the gamma ray yield as much as increasing
$\Delta x$.

We conclude that allowing the laser pulse to diverge over a distance of $80~\micron$
in a head-on collision is the optimal configuration to produce gamma rays. This can
be achieved experimentally by focussing the high-intensity laser pulse with an optic
that has an aperture in it. Provided this aperture is of sufficient size to permit
the transmission of the wakefield-driving laser, the electron beam and resultant gamma rays,
back-reflection and damage to the optical chain can be avoided.

An analytical fit to the region in which the total gamma ray yield is at least
80\% of its maximum value is
	\begin{eqnarray}
	\left( \frac{\Delta x - 91~\micron}{60~\micron} \right)^2
	+ \left( \frac{\Delta y}{3.6\micron} \right)^2 \leq 1
	\end{eqnarray}
for $\theta = 0\degree$. Even though the experiment would be designed for counterpropagation,
the angle between the beams will vary from shot to shot due to the pointing variation of
both the wakefield-driving and target laser pulses. \Fref{fig:ElectronBeam} shows that to achieve
any gamma ray production at $\Delta x = 80~\micron$, that angle must be less than $15\degree$.
We can understand this requirement by considering the
distance of closest approach between the centres of the electron beam and laser pulse
$\Delta_\mathrm{min} = \Delta x \sin \theta/2$; for $\Delta x = 80\micron$, $\theta > 15\degree$
means $\Delta_\mathrm{min} > 10~\micron$, the width of the electron beam, and so there is
no overlap between the beams.


\section{Energy loss of the electron beam}
\label{sec:ElectronBeam}

	\begin{figure}
	\centering
	\makeatletter
	\if@twocolumn
		\includegraphics[width=\linewidth]{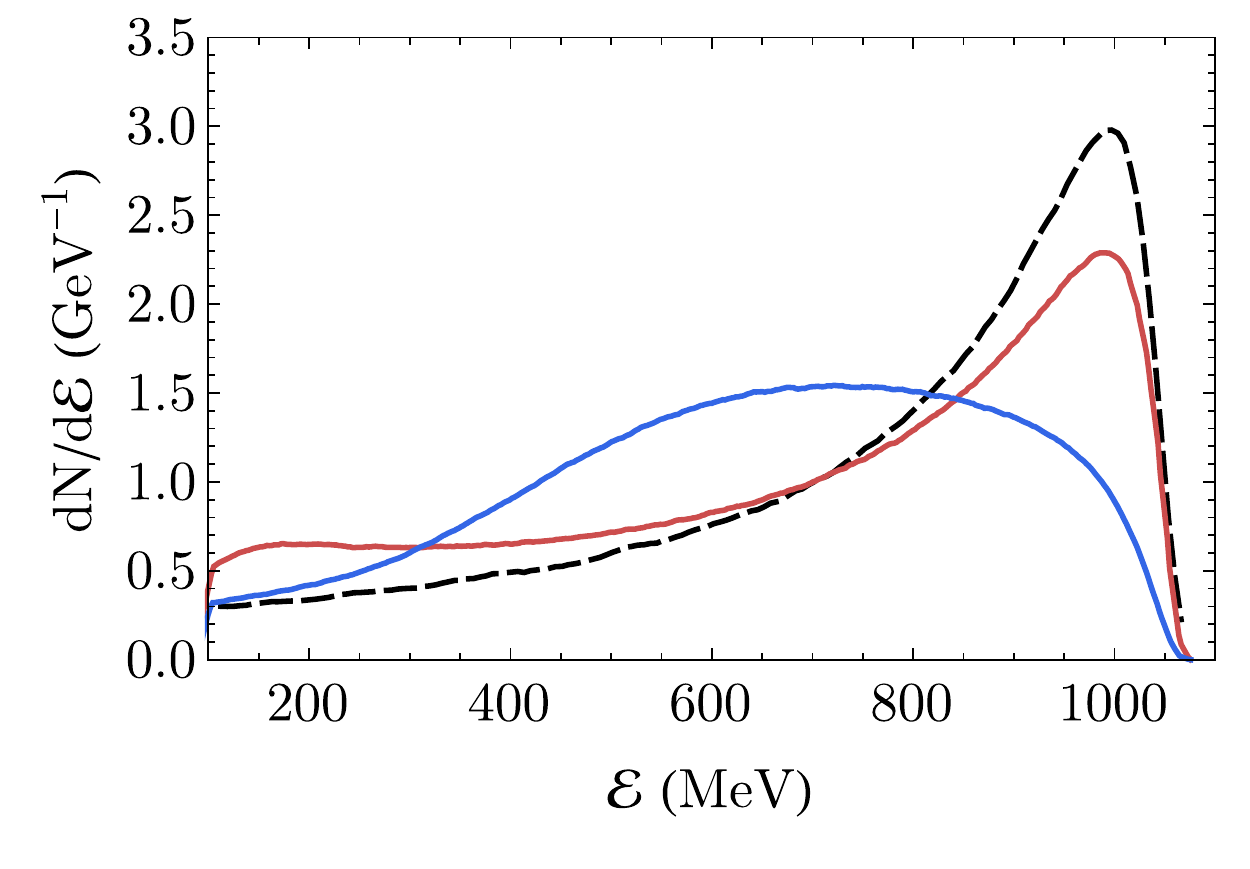}
	\else
		\includegraphics[width=0.6\linewidth]{figure4}
	\fi
	\makeatother
	\caption{The energy spectrum of the electron beam before the collision (black, dashed)
			and after a collision at $\Delta x = 0$ (red) and $90~\micron$ (blue).}
	\label{fig:ExamplePlot}
	\end{figure}

Detecting radiation reaction in an experiment could also be accomplished by comparing the
energy spectrum of a wakefield-accelerated electron beam with and without the target
short pulse, showing the reduced energy of the beam in the former case.
\Fref{fig:ExamplePlot} compares the energy distribution
of the GeV electron beam described in \sref{subsec:ElectronBeam} to its initial
energy spectrum for collisions at $\Delta x = 0$ and $90~\micron$.

We can see in that in the former case the spectra are not sufficiently
distinguishable, even though there is substantial loss of energy to gamma rays.
That loss is generated by those electrons that have
collided with the intense part of the laser pulse; however, the peak at 1000~MeV remains
as many electrons miss the laser pulse entirely. The broad energy spread is caused not only
by the range of peak intensities encountered by the electron beam but 
by the stochastic nature of emission as well: two electrons travelling along the same trajectory
will not necessarily lose the same energy.
Neitz and Di Piazza~\cite{neitz2013} showed that this leads to energy broadening even
for an electron beam without spatial extent and that radiation reaction manifests itself
by `smearing out' the initial energy distribution.

If we consider the final state energy distribution for a collision at
$\Delta x = 90~\micron$, enough of the beam interacts with the laser such that the
initial peak at 1000~MeV is entirely removed. If the electron beam is sufficiently
well-characterised, it may be possible to detect radiation reaction by comparing
the spectra from shots in the presence and absence of the target laser pulse. However, even
high-quality wakefield-accelerated electron beams do not in general have energy spectra
that are reproducible 
from shot to shot. An imperfectly generated electron beam would be a plausible origin
of the blue spectrum in \fref{fig:ExamplePlot}.

	\begin{figure}
	\centering
	\makeatletter
	\if@twocolumn
		\includegraphics[width=\linewidth]{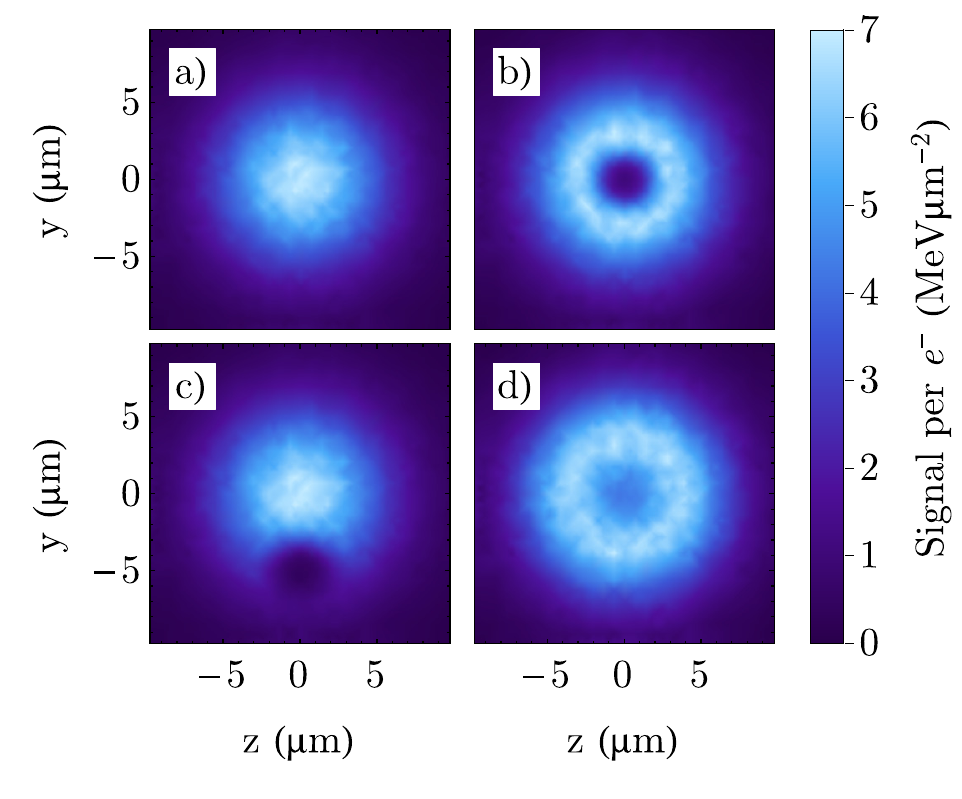}
	\else
		\includegraphics[width=0.6\linewidth]{figure5}
	\fi
	\makeatother
	\caption{The energy carried by the electron beam, per electron, per unit
			cross-sectional area a) prior to the collision and immediately after a collision at
			b) $\Delta x = \Delta y = 0$, c) $\Delta x = 0$, $\Delta y = 5~\micron$
			and d) $\Delta x = 50~\micron$, $\Delta y = 0$.}
	\label{fig:spectra}
	\end{figure}
	
\begin{figure}
	\centering
	\makeatletter
	\if@twocolumn
		\includegraphics[width=\linewidth]{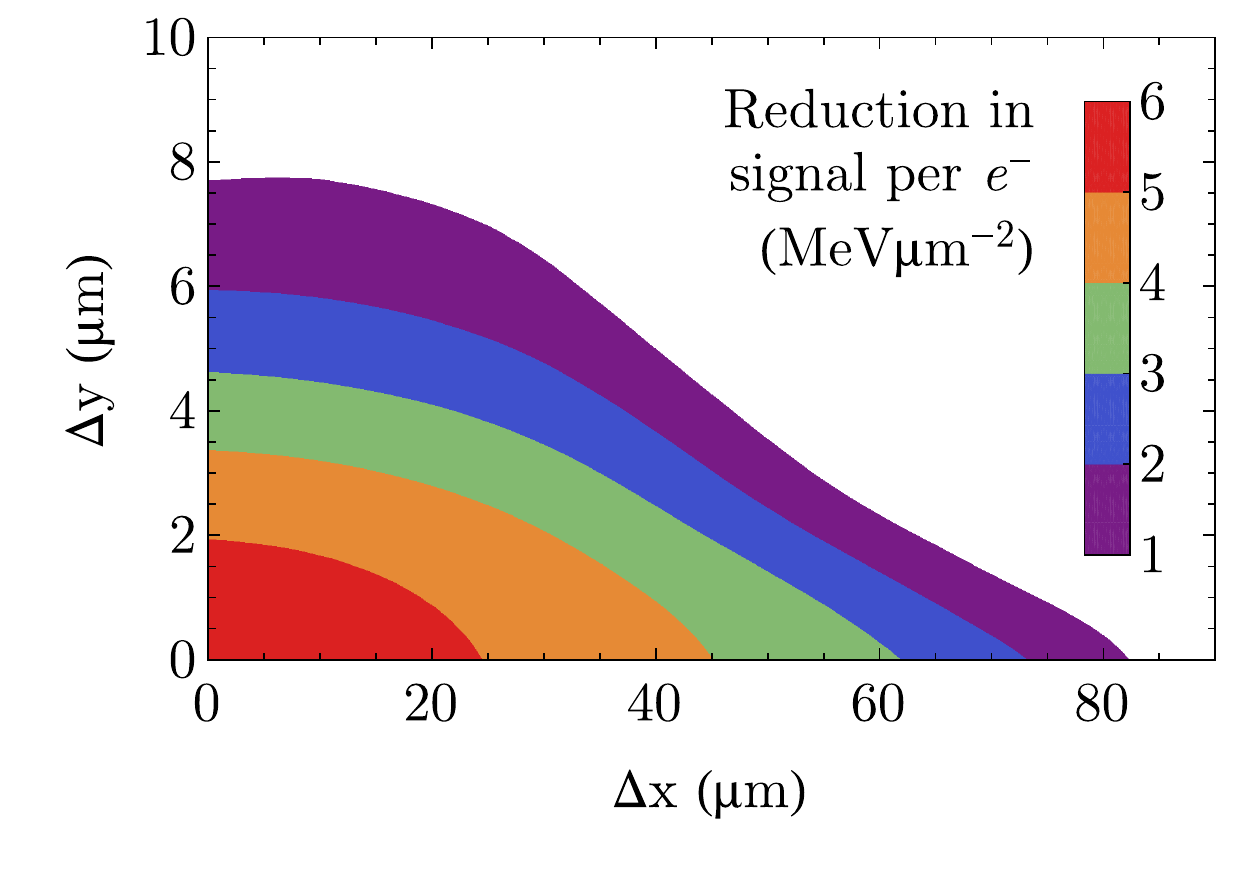}
	\else
		\includegraphics[width=0.6\linewidth]{figure6}
	\fi
	\makeatother
	\caption{The reduction in the energy carried by the electron beam per unit
			cross-sectional area, at the point where the laser pulse passes through
			the beam.}
	\label{fig:PeakQuality}
	\end{figure}

An alternative method of diagnosing radiation reaction exploits the fact that
the laser pulse has a smaller diameter
than the electron beam. \Fref{fig:PeakQuality} shows the energy
carried by the electron beam over its cross-sectional area, i.e. its energy
per unit volume integrated along its direction of propagation,
for electrons that have collided with the laser pulse for various $(\Delta x,
\Delta y, \theta)$. We can see by comparing a) and b) that the laser pulse causes significant
depletion of the energy spectrum in a region of radius $2~\micron$ around the
optical axis. Resolving the electron's beam areal energy spectrum could be
accomplished by allowing to diverge over a long distance
as it propagates away from the interaction region. A single shot would thereby allow
the simultaneous measurement of the electron beam energy in the presence
and absence of the target laser pulse and so the detection of radiation reaction.

If we introduce a perpendicular displacement between the beams $\Delta y = 5~\micron$,
as in c), the depletion zone is still evident. However, d) shows that a longitudinal
displacement, which allows the laser pulse to broaden before the collision takes place, makes the
depletion zone less obvious. This is because the laser interacts with more electrons but
at lower peak intensity, spreading the energy loss over a larger region of the electron beam.
By considering the amount by which the
energy is reduced at the centre of the depletion zone for collisions that
take place at $\theta = 0\degree$, as in \fref{fig:PeakQuality}, we can constrain the
accuracy of collision necessary for the depletion zone to be evident. An analytical
fit to the region in which that loss $> 3~\MeV \micron^{-2}$ is
	\begin{eqnarray}
	\frac{\Delta y}{4.6~\micron} + \left( \frac{\Delta x}{61~\micron} \right)^2 \leq 1.
	\end{eqnarray}


\section{Conclusions}

The gamma ray yield and the energy loss of the electron beam are
sensitive to both the maximal $\eta$ reached by the electrons as well as the length over which
this is sustained. However, for a realistic electron beam, the overlap between the beams will
be a more significant factor in determining the strength of radiation reaction.

Maximising these requires the counterpropagation of the electron beam and laser pulse;
this exploits the slow angular divergence of the laser pulse, minimising the accuracy
of longitudinal timing required, and the geometric factor $(1+\cos\theta)$ in $\eta$.

For a GeV electron beam colliding head-on with a 30~fs laser pulse of
intensity $10^{22}~\Wcm^{-2}$, the mean energy lost to
gamma rays will be maximised at 120~MeV per electron for a collision that occurs $80~\micron$ along the optical axis from the laser focal plane. If the collision occurs closer to the focal spot,
the gamma ray yield is reduced, but there will be a prominent depletion zone in the
areal energy spectrum of the electron beam.

The total energy emitted to gamma rays will be at least 80\% of its maximum if
the parallel and perpendicular offsets satisfy
	\begin{eqnarray}
	\left( \frac{\Delta x - 91~\micron}{60~\micron} \right)^2
	+ \left( \frac{\Delta y}{3.6\micron} \right)^2 \leq 1
	\label{eq:AccReq}
	\end{eqnarray}
and the depletion zone in the electron spectrum will be significant, i.e. the areal
energy will be reduced by more than $3~\MeV\micron^{-2}$ from its unperturbed value, if
	\begin{eqnarray}
	\frac{\Delta y}{4.6~\micron} + \left( \frac{\Delta x}{61~\micron} \right)^2 \leq 1.
	\label{eq:AccReq2}
	\end{eqnarray}
The latter allows the simultaneous measurement of the electron beam energy in the presence
and absence of the laser pulse and so the detection of radiation reaction.
	
A head-on collision can be accomplished by focussing the high-intensity target laser pulse
using an optic with an aperture of sufficient size to permit
the transmission of the wakefield-driving laser, the electron beam and resultant gamma rays.
Obtaining a $\Delta x$ and $\Delta y$ that satisfy
the accuracy requirements \eref{eq:AccReq} and \eref{eq:AccReq2} will rely on
gathering statistics over a large series
of laser shots. While the region in $\Delta x, \Delta y$ phase space is small,
the energy converted to gamma rays will be
sufficiently large, and the resulting depletion zone in the electron spectrum sufficiently
prominent, to be robust signatures of radiation reaction for experimental
parameters that can be achieved in current high intensity laser facilities.


\ack
This work was supported by an EPSRC studentship. The author thanks A.R. Bell and C.P. Ridgers
for continued advice and support.


\appendix
\section{The quantum synchrotron function}
\label{app:QSF}
\setcounter{section}{1}
The quantum synchrotron function is
	\makeatletter
	\if@twocolumn
	\begin{eqnarray}
	\fl F(\eta, \chi) =
		\frac{4 \chi}{3 \eta^2}
		&\left[
			\left(
				1 - \frac{2\chi}{\eta} + \frac{1}{1 - 2\chi/\eta}
			\right)
			K_{2/3} (\delta) \right. \nonumber
			\\
			&\quad\left.-
			\int_\delta^\infty \! K_{1/3} (t) \, \rmd t
		\, \right]
	\label{eq:QSF}
	\end{eqnarray}
	\else
	\begin{equation}
	\fl F(\eta, \chi) =
		\frac{4 \chi}{3 \eta^2}
		\left[
			\left(
				1 - \frac{2\chi}{\eta} + \frac{1}{1 - 2\chi/\eta}
			\right)
			K_{2/3} (\delta) -
			\int_\delta^\infty \! K_{1/3} (t) \, \rmd t
		\, \right]
	\label{eq:QSF}
	\end{equation}
	\fi
	\makeatother
where
	\begin{eqnarray}
	\delta = \frac{4 \chi}{3 \eta^2} \left( 1 - \frac{2\chi}{\eta} \right)^{-1}.
	\end{eqnarray}
For small $\chi$, this is approximately
	\begin{eqnarray}
	F(\eta, \chi) =
		\left(\frac{16}{3}\right)^{1/3}
		\Gamma\!\left(\frac{2}{3}\right)
		\eta^{-2/3} \chi^{1/3}.
	\end{eqnarray}
This means that $F(\eta, \chi)/\chi$, and so the differential rate of photon
emission~\eref{eq:MasterRate}, diverge as $\chi^{-2/3}$ for low frequencies.
However, the normalised energy emitted to photons with frequency between $\chi$
and $\chi + \rmd \chi$ during interval $\rmd t$,
$\chi \frac{\rmd^2 N}{\rmd t \rmd \chi} \rmd t \rmd \chi$, is well defined for all $\chi$.
In particular, it is zero for $\chi = 0$.
Similarly, the total rate of photon
emission~\eref{eq:TauRate}, which depends on $\int \! F(\eta, \chi)/\chi \, \rmd \chi$,
is always well defined.

\section{The typical $\eta$ of an electron emitting its first photon}
\label{app:PhaseDstr}

	\begin{figure*}
	\centering
	\includegraphics[width=0.45\linewidth]{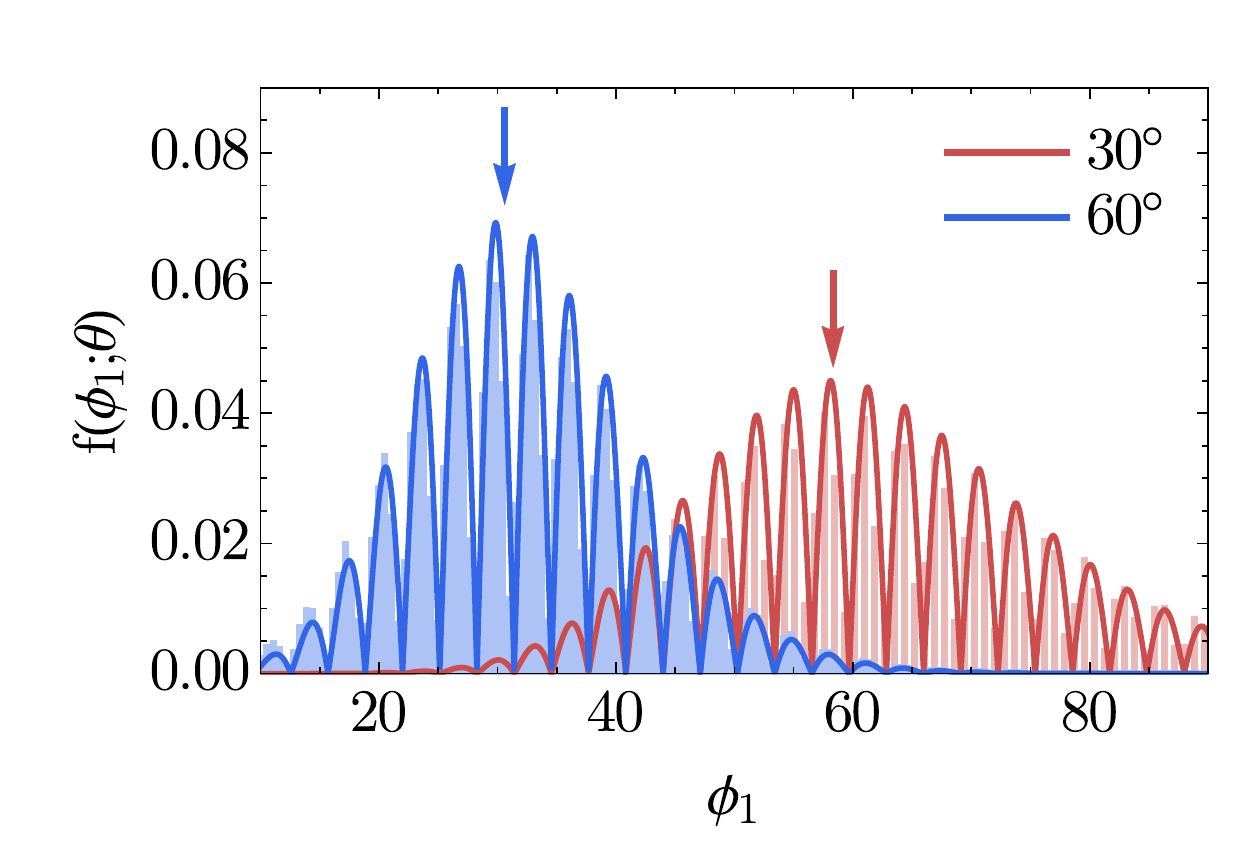}
	\includegraphics[width=0.45\linewidth]{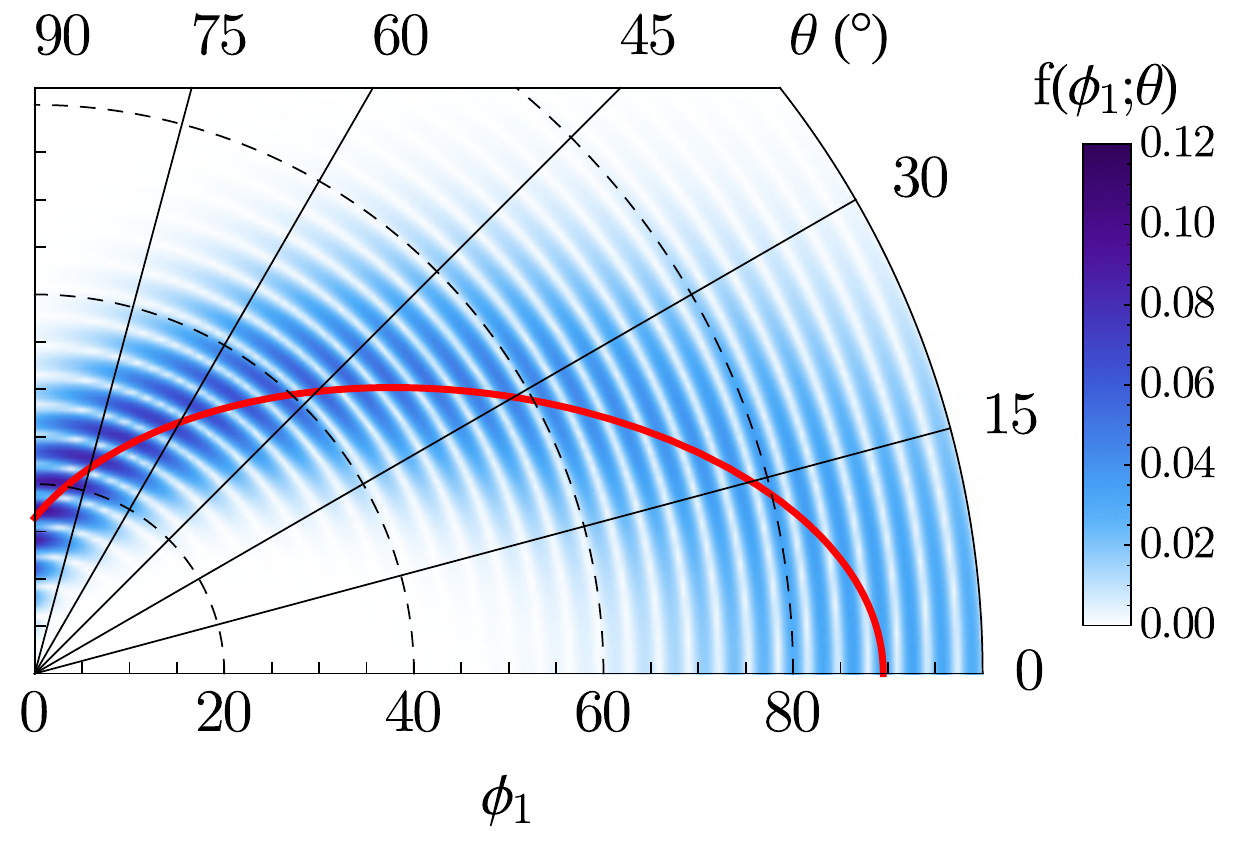}
	\caption{(left) The probability density that an electron colliding
			with Gaussian laser pulse of intensity
			$10^{22}~\Wcm^{-2}$
			at $\theta = 30\degree$ (red) and $60\degree$ (blue) emits its first
			photon at wave phase $\phi_1$. The curves are calculated using~\eref{eq:PhasePDF}
			and the vertical bars give the distribution found in simulations for a single
			electron with $\mathcal{E}_0 = 1000~\MeV$ colliding with a laser pulse as
			described in \sref{subsec:LaserPulse}.
			The arrows indicate the value
			of $\phi_{1,\mathrm{mp}}$ given by \eref{eq:ModalPhase}.
			(right) Density plot of $f(\phi_1;\theta)$: the red line is $\phi_{1,\mathrm{mp}}$.}
	\label{fig:PhaseDstrs}
	\end{figure*}

Recalling that
	\begin{eqnarray}
	\eta = \frac{\gamma |\mathbf{E}_\perp + \mathbf{v} \times \mathbf{B}|}{E_\mathrm{Sch}}
		= \frac {\gamma (1 + \cos \theta) |\mathbf{E}|}{E_\mathrm{Sch}}
	\end{eqnarray}
for an ultrarelativistic electron propagating at angle $\theta$ to a linearly
polarised electromagnetic wave, the rate of change of optical depth against emission is
	\begin{eqnarray}
	\frac{\rmd \tau}{\rmd t}
		= \frac{\sqrt{3}\alpha}{2 \pi \tau_\mathrm{C}}
			\frac{\eta}{\gamma} h (\eta)
		= \frac{\sqrt{3}\alpha}{2 \pi \tau_\mathrm{C}}
			(1 + \cos \theta)
			\frac{|E(t)|}{E_\mathrm{Sch}} h(\eta).
	\label{eq:OpticalDepthRate}
	\end{eqnarray}

The first approximation is that $h(\eta) \simeq h(0) = 5 \pi / 3$ at all
times along the electron trajectory. This is justified as in the foot of the pulse,
the electric field is sufficiently small to keep $\eta \lesssim 0.1$. Even then,
$h(0.1) = 0.93\,h(0)$ so any overestimate of the rate will be negligible.

In the presented simulations, $\mathcal{E}_0 = 1000~\MeV$ and therefore the electron can be treated
as ultrarelativistic. Therefore its initial momentum is
$\mathbf{p}_0 = -\mathcal{E}_0/c \, (\cos \theta, \sin \theta, 0)$.
As $\gamma_0 = \mathcal{E}_0 / m_e \gg a_0$, it can be assumed that any deflection of
the electron by the laser pulse is minimal. Its trajectory before emission is linear,
with position at time $t$ given by $(x,y,z) = -c t (\cos \theta, \sin \theta, 0)$.
As \eref{eq:OpticalDepthRate} depends on the electron energy only through $h(\eta)$,
which we have approximated as constant, the result will depend on $\mathcal{E}_0$ only in the sense
that it must be sufficiently large that the electron be ultrarelativistic.

In the simulations the full form of the Gaussian beam was used. However, for simplicity in
these calculations, neglecting wavefront curvature and beam divergence gives
	\begin{eqnarray}
	E &= E_0 \sin \phi \, \exp \!
		\left( -\frac{y^2+z^2}{w_0^2} - \frac{2\ln 2 \, \phi^2}{k^2 f^2} \right)
	\label{eq:Efld}
	\\
	\phi &= k x - \omega t = -(1 + \cos \theta) \omega t
	\label{eq:Phi}
	\end{eqnarray}
where $\omega$ is the laser's angular frequency, $k$ its wavevector, $w_0$ its waist size
and $f$ the full length at half maximum of its temporal intensity profile.

Using \eref{eq:Phi} to recast \eref{eq:OpticalDepthRate} and \eref{eq:Efld} in terms of
$\phi$, we find
	\begin{eqnarray}
	\frac{\rmd \tau}{\rmd \phi} =
		-\frac{5\alpha}{2 \sqrt{3} \omega \tau_\mathrm{C}}
		\frac{E_0}{E_\mathrm{Sch}}
		|\sin \phi|
		\, \exp \! \left( - \zeta^2 \phi^2 \right)
	\label{eq:TauPhi}
	\end{eqnarray}
where
	\begin{eqnarray}
	\zeta^2 =
		\frac{\sin^2 \theta}{(1 + \cos \theta)^2 k^2 w_0^2} + \frac{2\ln 2}{k^2 f^2}.
	\label{eq:Zeta}
	\end{eqnarray}

As the envelope in \eref{eq:TauPhi} varies slowly with $\phi$, we can approximate
the oscillatory factor $|\sin \phi| \simeq 2 / \pi$ and thus integrate \eref{eq:TauPhi}
from $\infty$ to $\phi_1$, the wave phase at which the electron emits its first photon.
We find the following relation between the final optical depth and $\phi_{1}$:
	\begin{eqnarray}
	\tau_\mathrm{f} \simeq
		\frac{5\alpha}{2 \sqrt{3 \pi} \omega \tau_\mathrm{C}}
		\frac{E_0}{E_\mathrm{Sch}}
		\frac{1 - \mathrm{erf} \left( \zeta \phi_1 \right)}{\zeta}.
	\label{eq:Tau}
	\end{eqnarray}

Photon emission is a stochastic process and governed by Poisson statistics, so these final
optical depths are distributed as $\tau_\mathrm{f} \sim \exp (-\tau_\mathrm{f})$. Then
the probability density that an electron colliding with a Gaussian laser pulse at angle
$\theta$ emits its first photon at phase $\phi_1$ is given by
	\begin{eqnarray}
	f(\phi_1; \theta) = \exp(-\tau_\mathrm{f}) \left| \frac{\rmd \tau_\mathrm{f}}{\rmd \phi} \right|
	\label{eq:PhasePDF}
	\end{eqnarray}
where we substitute \eref{eq:Tau} in the exponent and \eref{eq:TauPhi} in the modulus.
\Fref{fig:PhaseDstrs} compares the analytical $f(\phi_1; \theta)$ with that obtained
from simulation and finds good agreement.

It is evident from \fref{fig:PhaseDstrs} that at larger angles of collision, the
electron is more likely to penetrate further into the laser pulse before emitting its
first photon; both the peak and the width of the distribution decrease as $\theta$ grows,
reducing the length of interaction. The most probable $\phi_1 = \phi_{1,\mathrm{mp}}$ can
be estimated by solving $\rmd f(\phi_1;\theta)/\rmd \phi_1 = 0$, using \eref{eq:Tau}, to find
	\begin{eqnarray}
	\phi_{1,\mathrm{mp}}^2 \simeq
		\frac{1}{2 \zeta^2}
		\, W \!
		\left[
			\frac{1}{\zeta^2}
			\left(
			\frac{5\alpha}{\sqrt{6}\pi} \frac{1}{\omega\tau_\mathrm{C}}
			\frac{E_0}{E_\mathrm{Sch}}
			\right)^2
		\right]
	\label{eq:ModalPhase}
	\end{eqnarray}
where $W(x)$ is the Lambert $W$ function, defined by $x = W(x) e^{W(x)}$, and $\zeta^2$ is as
given in \eref{eq:Zeta}.

We can substitute \eref{eq:ModalPhase} into $|\mathbf{E}| \simeq E_0 \exp (-\zeta^2 \phi^2)$
and then $\eta = \gamma (1+\cos\theta) |\mathbf{E}|/E_\mathrm{Sch}$ to arrive at an estimate
of the typical $\eta$ of an electron emitting its first photon:
	\begin{eqnarray}
	\eta \simeq
		\frac{\gamma_0 (1 + \cos \theta) E_0}{E_\mathrm{Sch}}
		\exp (-\zeta^2 \phi_{1,\mathrm{mp}}^2).
	\label{eq:TypicalEta}
	\end{eqnarray}
For a GeV electron colliding with the laser pulse described in \sref{subsec:LaserPulse}, it
is maximised for $\theta = 65\degree$.


\end{document}